\documentclass[preprint,10pt]{aastex}

\usepackage{amsmath,amsthm,amsfonts,amssymb,relsize}
\usepackage{graphicx,array,MnSymbol,bbold,hyperref}
\usepackage[usenames,dvipsnames]{xcolor}

\theoremstyle{definition}

\theoremstyle{remark}

\newcommand{\lp}{\left(}  
\newcommand{\rp}{\right)} 
\newcommand{\lb}{\left[}   
\newcommand{\rb}{\right]}  

\newcommand{\mb}{\mathbf} 
\newcommand {\cl}{\mathcal} 
\newcommand {\tsf} [1]{\textsf{#1}} 
\newcommand{\vectsym}{\boldsymbol}  
\DeclareMathOperator{\Tr}{Tr\,}            
\DeclareMathOperator{\sinc}{sinc\,}       

\newcommand{\comm}[2] {\left[#1,#2\right]}

\newcommand {\ket}[1] {\left|{#1}\right\rangle}

\newcommand {\bra}[1] {\langle{#1}|}

\newcommand{\abs}[1]{\left | #1 \right |}   
\newcommand{\mean}[1]{\left\langle #1 \right\rangle}  

\def\undertilde#1{\mathord{\vtop{\ialign{##\crcr
$\hfil\displaystyle{#1}\hfil$\crcr\noalign{\kern1.5pt\nointerlineskip}
$\hfil\tilde{}\hfil$\crcr\noalign{\kern1.5pt}}}}}

\begin{document}
\title{Quantum optimality of photon counting for temperature measurement of thermal astronomical sources}
\author{Ranjith Nair$^{1}$ and Mankei Tsang$^{1,2}$}
\affil{${}^{1}$ Department of Electrical and Computer Engineering, 4 Engineering Drive 3, \\National University of Singapore, Singapore 117583\\
${}^{2}$ Department of Physics,  2 Science Drive 3,\\
 National University of Singapore, Singapore 117551}
\begin{abstract} 
Using the quantum Cram\'{e}r-Rao bound from quantum estimation theory, we derive a fundamental quantum limit on the sensitivity of a temperature measurement of a thermal astronomical source. This limit is expressed in terms of the source temperature $T_s$, input spectral bandwidth $\Delta \nu$, and measurement duration $T$, subject  to a long measurement time assumption $T\Delta\nu \gg1$. It is valid for any measurement procedure that yields an unbiased estimate of the source temperature. The limit agrees with the sensitivity of direct detection or photon counting, and also with that of the ideal radiometer in the regime $kT_s/h\nu_0 \gg 1$ for which the Rayleigh-Jeans approximation is valid, where  $\nu_0$ is the center frequency at which the radiometer operates. While valid across the electromagnetic spectrum, the limit is especially relevant for radio astronomy in this regime, since it implies that no ingenious design or technological improvement can beat an ideal radiometer for temperature measurement. In this connection, our result refutes the recent claim of a radio astronomy technique with much-improved sensitivity over the radiometer \citep{LKD15}.
\end{abstract}
\maketitle
\section{Introduction}

Astronomical observations are made over frequencies ranging from radio {and}  microwave frequencies through to infrared, optical, X-ray, and gamma-ray frequencies. However, measurement techniques and instruments vary widely over the electromagnetic spectrum. At radio to millimeter-wave frequencies, the chief measurement device in astronomy is the radiometer \citep{Dic46}. For thermal sources with photon flux spectral density $n_0 \;\mbox{(in photons}\cdot\mbox{s}^{-1}\cdot\mbox{Hz}^{-1})$, related to the source temperature $T_s$, and bandwidth $\Delta \nu$ Hz, the sensitivity of an ideal radiometric measurement of $n_0$ follows the radiometer equation
\begin{align} \label{radiometer}
\frac{\mbox{Var}\,\hat{n}_0^{\mbox{\relsize{-2}(rad)}}}{n_0^2} = \frac{1}{\Delta\nu T},
\end{align}
where the left-hand side is the \emph{relative sensitivity} $\mbox{(Variance / squared mean)}$ of the radiometer estimate $\hat{n}_0^{\mbox{\relsize{-2}(rad)}}$ of $n_0$ and $T$ is the measurement time \citep{BG-S10}.
Recently, \citet{LKD15}  proposed a two-detector setup similar to the intensity interferometer of Hanbury Brown and Twiss \citep{HBT57} and an estimator $\hat{n}_0^{\mbox{\relsize{-2}(LKD)}}$ of the source temperature that was claimed to achieve the relative sensitivity
\begin{align} \label{LKDresult}
\frac{\mbox{Var}\,\hat{n}_0^{\mbox{\relsize{-2}(LKD)}}}{n_0^2} \overset{?}{=} 5\frac{T_{\tsf{samp}}}{T} + \frac{1}{n_0 \Delta\nu T}.
\end{align}
This equation is essentially the square of eqn.~(33) of \citep{LKD15}, with the following relabeling for consistency with the rest of this paper. We have renamed  $T$ of eq.~(33) of \citep{LKD15} to $T_{\tsf{samp}}$ -- the time needed to obtain one sample in their scheme. Their number of samples $N$ is then $N = T/T_{\tsf{samp}}$, where here -- and throughout this paper -- $T$ stands for the \emph{total} observation time needed to produce all samples. We have also replaced their coherence time $\tau$ with the inverse of the bandwidth $\Delta \nu$, which correspondence holds to within a constant numerical factor for most thermal spectra of interest.  The sensitivity of eq.~\eqref{LKDresult} is claimed under the condition $T_{\tsf{samp}} \ll 1/\Delta \nu$, requiring fast, but not unfeasibly fast, detection electronics. In principle however, $T_{\tsf{samp}}$ can be made arbitrarily small relative to $1/\Delta \nu \ll T$, so that the second term is the limiting one fundamentally. Since $n_0 \Delta \nu T$ is the average number of photons incident during the observation period, eqn.~\eqref{LKDresult} shows a  \emph{Poisson} scaling of the relative sensitivity characteristic of coherent-state sources. In particular, the sensitivity decreases (i.e., the measurement improves) with increasing source temperature, whereas the radiometer equation \eqref{radiometer} exhibits no dependence of the sensitivity on $n_0$. In the photon-rich RF regime, the scaling of eq.~\eqref{LKDresult}, if correct,  can offer orders-of-magnitude improvements in sensitivity.

In \citep{Zmu15}, Zmuidzinas has examined in detail the derivation of eq.~\eqref{LKDresult}, and concluded that the two-detector scheme of \citet{LKD15} actually leads to a relative sensitivity
\begin{align} \label{Zmuresult}
\frac{\mbox{Var}\,\hat{n}_0^{\mbox{\relsize{-2}(LKD)}}}{n_0^2} =  \frac{n_0 + 1}{n_0 \Delta\nu T},
\end{align}
(cf. eqns.~(136) and (E21) of \citep{Zmu15}) which is in close agreement with the radiometer equation. Moreover, Zmuidzinas has shown that simple photon counting (direct detection) -- to which the radiometer provides a close approximation at radio frequencies  for photon-rich thermal sources \citep{Nit94,Zmu03} -- gives an identical sensitivity. In other words, the scheme of \citep{LKD15} is no better than usual radiometry and does not provide the purported increased sensitivity. We refer the reader to \citep{Zmu15} for extensive discussion on the logical fallacy leading to the incorrect result eq.~\eqref{LKDresult} for the two-detector scheme, and the physical reasons why that scheme cannot work as claimed.

Our focus in this paper is different. Insofar as the radiation incident on a telescope of any kind, irrespective of the wavelength region, is electromagnetic in nature, it is governed by the quantum theory of radiation. Any measurement scheme aimed at extracting information about a parameter  such as $n_0$ from the incident field is subject to fundamental limitations following from the laws of quantum mechanics. The study of the limitations on extracting information from quantum systems using quantum measurements is the subject matter of \emph{quantum estimation theory} \citep{Hel76,Hol11}, which falls under the general rubric of quantum metrology \citep{GLM11}  with close connections to the field of quantum information \citep{NC00}. 

Using the ideas of quantum estimation theory, in particular the quantum Cram\'{e}r-Rao bound, we show that any method of estimating the noise temperature of a thermal source suffers from a \emph{minimum} relative sensitivity equal to the right-hand side of eq.~\eqref{Zmuresult} in the limit of large observation times $T$ relative to the coherence time $\tau_c = 1/\Delta \nu$ of the incident radiation. This result provides another refutation, independent of the work of \citet{Zmu15}, of the claim of eq.~\eqref{LKDresult}. More importantly, however, it shows that the relative uncertainty of eq.~\eqref{Zmuresult} cannot be beaten by \emph{any conceivable measurement scheme}, provided only that the scheme yields an unbiased estimate in the sense  that the average of many estimates of the parameter converges to the true value  -- we note that both photon counting and the scheme of \citep{LKD15} lead to unbiased estimates and therefore fall under the purview of the bound.

{ We briefly mention some recent work on quantum limits for temperature measurement in different contexts from ours.  \citet{MB13,MB15} have studied limits on the accuracy of temperature measurement of quantum gases using the quantum Cram\'{e}r-Rao bound and found that energy measurements are optimal.  \cite{JZ14} have studied the local estimation accuracy  -- i.e., the accuracy of estimating small deviations of temperature from a preset fiducial value -- of pyrometers, i.e., devices that measure the total energy output  of a thermal source. Under this assumption, they find that the accuracy of local estimation of temperature matches that from the quantum Cram\'{e}r-Rao bound. In this paper, the temperature range of the source is not restricted and explicit account is also taken of the spatial and temporal bandwidth constraints relevant to astronomical observations.}

This paper is organized as follows. In Section 2, we first review the quantum theory of thermal radiation. We use a full multi-temporal mode analysis to derive the density operator of filtered thermal radiation in the limit of coherence time $\tau_c \ll T$, the observation time.  This condition is typically satisfied for both radio and optical astronomical observations. { The details of the calculation appear in the Appendix.} In Section 3, we briefly review the quantum Cram\'{e}r-Rao bound before applying the results of Section 2 to explicitly evaluate it for filtered thermal radiation. We obtain the result that  the right-hand side of eq.~\eqref{Zmuresult} is a fundamental quantum limit to the relative sensitivity of any unbiased estimator of the source temperature. We close with a discussion of the implications of this result in Section 4.

\section{Quantum-mechanical description of input field}

\subsection{Filtered thermal radiation}
Regardless of the frequency range, the electromagnetic field input to an antenna or telescope is described by a time-dependent positive-frequency field operator ${E}(t)$ (in units of  $\sqrt{\mbox{photons}\cdot\mbox{s}^{-1}}$) given by
\begin{align}
{E}(t) = \int_{0}^{\infty}  a_{\nu} \,e^{-i 2\pi \nu t}\,d{\nu}.
\end{align}
Here, the $\{a_{\nu}\}_{\nu>0}$ are un-normalized single-frequency annihilation operators satisfying $\comm{a_{\nu}}{a^{\dag}_{\nu'} }= \delta(\nu - \nu')$ which implies that $\comm{E(t)}{E^{\dag}(t')} = \delta(t-t')$. As in \citep{LKD15,Zmu15}, we are assuming the input field to be of a single polarization and in a single spatial mode to focus on the central issue -- the additional generality does not substantially alter the result. 
If the input field is from a thermal source, it is a \emph{Gaussian field} \citep[][Sec.~III]{Sha09} with the quantum expectation values \citep{MW95,Zmu03}:-
\begin{align}
\mean{a_\nu} &=0,  \\
\mean{a_{\nu}\, a_{\nu'}} &=0 \\
\mean{a^\dag_{\nu'}\, a_{\nu}} &= n_{\tsf{th}}(\nu)\,\delta(\nu-\nu'), \label{occnumold}
\end{align}
where $n_{\tsf{th}}(\nu)$ is the mean occupation number in a thermal state given by the Planck formula
\begin{align} \label{planck}
n_{\tsf{th}}(\nu) = \frac{1}{e^{\frac{h\nu}{kT_s}}-1},
\end{align}
for $T_s$ the source temperature (in K) and $k$, Boltzmann's constant. In terms of field operators, these relations imply 
\begin{align} 
\mean{E(t)} &= 0, \label{mean} \\
K^{(p)}(t,t') &:= \mean{E(t)\,E(t')} = 0,  \label{Kp}\\
\label{Knold}
K^{(n)}(t,t') &:= \mean{E^\dag(t)\, E(t')} = \int_{0}^{\infty} n_{\tsf{th}}(\nu)\, e^{-i2\pi \nu (t'-t)}\, d{\nu}.
\end{align}
Here, the functions $K^{(p)}(t,t')$ and $K^{(n)}(t,t')$ are the phase-sensitive and (normally ordered) phase-insensitive correlation functions of the field  respectively (cf. eqns.~(61)-(62) of \citet{Sha09}). Note that these functions (as well as the mean) depend only upon the time difference $\tau := t'-t$, indicating the statistical  stationarity of the field. Using the Gaussian moment-factoring theorem \citep{MW95}, all higher  moments of the field operators can be expressed in terms of these functions, which therefore constitute a complete description of the field.

The thermal field described above is broadband. In practice, a measurement operates on only a finite band of the input field that is determined either by insertion of filters or the response of the measuring device. We will accordingly assume that the input is passed through a filter centered at frequency $\nu_0$ and with a flat profile over the band $[\nu_0 - \Delta\nu/2, \nu_0 + \Delta \nu/2]$. In a typical radio astronomy measurement, e.g., we may have $\nu_0 = 1$ GHz, and $\Delta \nu$ a few MHz. Further, in radio astronomy, the source temperature $T_s$ is typically such that  $h\nu_0/kT_s \ll 1$, so that the Rayleigh-Jeans approximation to eq.~\eqref{planck} may be used \citep{BG-S10}. As a further simplification, we may assume that the mean occupation number of eq.~\eqref{planck} is approximately flat in the band $[\nu_0 - \Delta\nu/2, \nu_0 + \Delta \nu/2]$ at the value $n_0 \simeq\ kT_s/h\nu_0$. As a result of the above assumptions, for the field at the output of the filter, we replace $n_{\tsf{th}}(\nu)$ in eqns.~\eqref{occnumold} and \eqref{Knold} by
\begin{align} \label{psd}
n(\nu)=
\left\{
	\begin{array}{ll}
		n_0 = \frac{k T_s}{h\nu_0} & \mbox{if } \nu \in[ \nu_0 - \Delta\nu/2,\nu_0 + \Delta\nu/2]  \\
		0 & \mbox{otherwise}.
	\end{array}
\right.
\end{align}
In this quasi-monochromatic regime, $n(\nu)$ is essentially the (dimensionless) power spectral density of the field, in that multiplication by $h\nu_0$ gives the average power per unit frequency  (in W$\cdot \mbox{Hz}^{-1}$) of the field. 

The field at the output of the bandpass filter continues to satisfy eqns.~\eqref{mean} and \eqref{Kp} but Eqn.~\eqref{Knold} is modified to
\begin{align} \label{Kn}
K^{(n)}(t,t') \equiv K^{(n)}(\tau) =  n_0\cdot \Delta\nu\, \sinc\lb\Delta\nu \tau \rb\, e^{-i2\pi \nu_0 \tau},
\end{align}
where $\sinc(x) = \sin(\pi x)/(\pi\,x)$ is the sinc function. The coherence time $\tau_c$ of the output radiation \citep{Man59}, which is  the approximate ``width'' of  $\abs{K^{(n)}(\tau)}$, is then
\begin{align} 
\tau_c := \int_{-\infty}^\infty \abs{g^{(1)}(\tau)}^2\, d\tau = \int_{-\infty}^\infty \abs{\frac{K^{(n)}(\tau)}{K^{(n)}(0)}}^2\, d\tau = \frac{1}{\Delta \nu},
\end{align}
and is also the separation between the peak of $\abs{K^{(n)}(\tau)}$ at $\tau=0$ and the first zero. For the radio frequency example above, we have $\tau_c \sim 1\, \mu$s. In the optical regime, $\tau_c$ is typically much smaller even downstream of an optical filter, of the order of nanoseconds. 

The measurement on the field takes place in a finite time interval $\cl{T}$ of duration $T\,$s, which we take to be $ [-T/2,T/2]$. In the remainder of the paper, we assume that the parameters $\nu_0, \Delta \nu$ (or $\tau_c$), and $T$ are fixed and known, while $n_0$ (or equivalently, $T_s$) is the single unknown parameter that we wish to estimate. We further assume that $\tau_c \ll T$, which is usually the case for the detection of faint sources, for which  $T$ could {range from seconds to hours to days.}

\subsection{Modal description}
Quantum information and metrology using electromagnetic fields \citep{Hel76,Hol11}, and Gaussian quantum information in particular \citep{WPG+12,Oli12}, is usually set up and studied on a finite set of modes that are excited in the problem under consideration.   In order to obtain a modal description of the field from that in terms of its mean and correlation functions in the previous subsection, we expand the field in the measurement interval $\cl{T}$ in terms of a complete orthonormal set of positive-frequency ``Fourier-series'' traveling-wave modes given by
\begin{align} \label{modefuncs}
\phi_m(t) = \left\{
	\begin{array}{ll}
		\frac{1}{\sqrt{T}} \exp\lp-i \frac{2\pi mt}{T}\rp  & \mbox{if } t \in[-T/2,T/2]  \\
		0 & \mbox{otherwise,}
	\end{array}
\right.
\end{align}
for $m=0,1,2,\ldots$. The modal annihilation operators
\begin{align} \label{annop}
a_m := \int_{-\infty}^{\infty} E(t) \phi^*_m(t)\, dt, \;\;\;\;m = 0,1,2,\ldots
\end{align}
then satisfy the (normalized) commutation relations
\begin{align}
\comm{a_m}{a_n^\dag} = \delta_{mn}.
\end{align}
As in standard quantum mechanics, the field state is described by a density operator $\rho$ (a positive semidefinite operator with $\Tr \rho =1$) on a Hilbert space $\cl{H} = \bigotimes_{m=0}^{\infty} \cl{H}_m,$ where $\cl{H}_m$ is the (infinite-dimensional) Hilbert space of the $m$-th mode. Our next task is to obtain $\rho$.

Since $E(t)$ is a Gaussian field, the density operator $\rho$ is in a so-called \emph{Gaussian state} \citep{Sha09,Hol11}. In order to define Gaussian states, we need some notation.
For each $m$, we define the quadrature operators
\begin{align} \label{quadops}
q_m = \frac{a_m + a^\dag_m}{\sqrt{2}};\;\; p_m =  \frac{a_m - a^\dag_m}{\sqrt{2}i}
\end{align}
satisfying the canonical commutation relations $[q_m, q_n] = [p_m, p_n]=0$ and $\comm{q_m}{p_n} = i\delta_{mn}$.
Consider the vector  $\mb{R} \equiv (R_1, R_2,\ldots) := (q_1, p_1, q_2, p_2, \ldots)$ of quadrature operators. The \emph{mean vector} $\overline{\mb{R}}$ in the state $\rho$ is
\begin{align}
\overline{\mb{R}} := \mean{\mb{R}}_\rho = (\Tr(\rho R_1), \Tr(\rho R_2),  \ldots)
\end{align}
and the \emph{covariance matrix} $\vectsym{\sigma}$ has the $(i,j)$-th matrix element 
\begin{align} \label{cov}
\sigma_{ij} := \frac{1}{2}\langle{(R_i - \overline{R}_i)\,(R_j - \overline{R}_j) +(R_j - \overline{R}_j)\,(R_i - \overline{R}_i)}\rangle_\rho = \mbox{Re}\left\{ \Tr \rho \,(R_i - \overline{R}_i)\,(R_j - \overline{R}_j) \right\}.
\end{align}
With this notation, a \emph{Gaussian state} is a state whose Wigner characteristic function 
\begin{align}
\chi_\rho(\vectsym{\xi}) := \Tr \lb \rho\, \exp \lp - \vectsym{\xi}^T\, \vectsym{\Omega}\, \mb{R}\rp \rb
\end{align} 
is of the Gaussian form \citep{Oli12}
\begin{align} \label{GCF}
\chi_\rho\lp \vectsym{\xi}\rp = \exp \lp - i \vectsym{\xi}^T\, \vectsym{\Omega}\, \overline{\mb{R}}  - \frac{1}{2} \vectsym{\xi}^T \vectsym{\Omega} \vectsym{\sigma} \vectsym{\Omega}^T\vectsym{\xi}\rp,
\end{align}
where $\vectsym{\xi} = (\xi^{(1)}_1, \xi^{(1)}_2, \xi^{(2)}_1, \xi^{(2)}_2, \ldots)$ is the vector of Fourier variables corresponding to the phase-space coordinates of each mode. The matrix 
\begin{align}
\vectsym{\Omega} = \bigoplus_{m} \vectsym{\omega}
\end{align}
is block-diagonal in the $2 \times 2$ blocks
\begin{align}
\vectsym{\omega} = \begin{pmatrix} 0 & 1 \\ -1 & 0 \end{pmatrix},
\end{align}
with one block per mode.

From eq.~\eqref{GCF}, we see that a Gaussian state is completely described by its mean and covariance matrix.  To calculate these for filtered thermal radiation, we must compute the first- and second-order moments of the form $\mean{a_m}$, $\mean{a_m\,a_n}$, and $\mean{a_m^{\dag}\,a_n}$ for each $m$ and $n$. { In the Appendix, we present the detailed calculations for the above quantities in the long observation time limit $\tau_c \ll T$. The results are
\begin{align}
\mean{a_m} & =0 \\
\mean{a_m\,a_n} &=0 \\
\mean{a_m^{\dag}\,a_n}&\simeq n_0\, \mbox{rect}\lb \frac{\nu_0 - \frac{m}{T}}{\Delta \nu} \rb\cdot \delta_{mn},
\end{align}
where 
\begin{align}
\mbox{rect} (x)=
\left\{
	\begin{array}{ll}
		1  & \mbox{if }\abs{x}\leq 1/2  \\
		0 & \mbox{otherwise}
	\end{array}
\right.
\end{align}
is the rectangle function.
}
In other words, in this $\tau_c \ll T$ limit, we have $\mean{a^{\dag}_m\,a_n} \simeq 0$ for $m \neq n$ and the average number of photons
\begin{align} \label{apn}
\mean{a^{\dag}_m\,a_m} \simeq n_0\;\;\mbox{for}\;\; m \in \cl{M} = \left\{\lp\nu_0 - \frac{\Delta\nu}{2}\rp T, \lp\nu_0 - \frac{\Delta\nu}{2}\rp T+1, \cdots, \lp\nu_0 + \frac{\Delta\nu}{2}\rp T\right\}
\end{align}
i.e., in $M:= T \Delta\nu$  ``approximately single-frequency'' modes that are within the bandwidth of the filter. 
Together with eqs.~\eqref{p}-\eqref{q}, eq.~\eqref{apn} implies that the covariance matrix $\vectsym{\sigma}$ of $\rho$ is
\begin{align}
\vectsym{\sigma} =  \mathlarger{\bigoplus_{m \in \cl{M}}} \lp \frac{2n_0 +1}{2}\, \mathbb{1}_2 \rp\,  \mathlarger{\bigoplus_{m \notin \cl{M}}} \lp\mathbb{1}_2 \rp,
\end{align}
where $\mathbb{1}_2$ is the $2 \times 2$ identity matrix. This implies that $M = \Delta \nu T$ of the modes (those in the set $\cl{M}$ defined in eq.~\eqref{apn}) are in independent thermal states each with $n_0$ photons on average given by \citep{Oli12}:-\begin{align} \label{thstate}
\rho_{\tsf{th}}(n_0) = \frac{1}{n_0 +1} \sum_{k=0}^\infty \lp \frac{n_0}{n_0+1} \rp^k \ket{k} \bra{k} = \frac{1}{\pi \,n_0} \int_{\mathbb{C}} \exp \lp - \frac{\abs{\alpha}^2}{n_0}\rp \ket{\alpha} \bra{\alpha}\, d^2\alpha.
\end{align}
In the first representation, $\rho_{\tsf{th}}(n_0)$ is a mixture of number states with a Bose-Einstein distribution, while in the second, it is a zero-mean circularly-symmetric Gaussian distribution of coherent states $\{\ket{\alpha}\}$, i.e., of the eigenstates of the annihilation operator -- $a \ket{\alpha} = \alpha\, \ket{\alpha}$ -- of the relevant mode. The remaining modes are all in the vacuum state $\ket{0}\bra{0}$. Thus, the overall state is
\begin{align} \label{state}
\rho_{n_0} = \lp \bigotimes_{m \in \cl{M}} \rho_{\tsf{th}}^{(m)}(n_0)\rp \lp \bigotimes_{m \notin \cl{M}} \ket{0}^{(m)(m)}\bra{0}\rp 
\end{align}
Since the modes $m \notin \cl{M}$  carry no information about $n_0$, it is sufficient to make measurements on just the modes in the set $\cl{M}$, effectively reducing the problem to one involving a finite number of modes.

\section{The quantum limit on estimating source temperature}

\subsection{The Quantum Cram\'{e}r-Rao bound}
The \emph{classical} Cram\'{e}r-Rao bound \citep{Cra46,Rao45,VanTreesI} provides a lower bound on the variance $\mbox{Var} \;\hat{\theta}$ of any unbiased estimator $\hat{\theta}$ of an unknown parameter $\theta$ indexing a family of probability distributions $\{P_\theta\}$ on a given sample space.  An \emph{unbiased} estimator of $\theta$ is one that satisfies $\tsf{E}[\hat{\theta}]=\theta$, where the statistical expectation value is taken with respect to $P_\theta$. The Cram\'{e}r-Rao bound is widely used to provide limits on and benchmarks for the performance of communication and measurement systems \citep{VanTreesI}. It has also found application in the design of astronomical instruments \citep{Zmu03b} and in measurements of the cosmic microwave background \citep{YKW07}. 

In a \emph{quantum} estimation problem, instead of probability distributions $\{P_\theta\}$, we are provided with a family  $\{\rho_\theta\}$ of density operators of a given quantum-mechanical system depending on the unknown parameter. The additional feature of the quantum estimation problem over its classical counterpart is the freedom of choosing the \emph{quantum measurement} that generates a probability distribution from $\rho_\theta$. All possible quantum measurements can be mathematically described by an object called a \emph{positive-operator-valued measure  (POVM)} \citep{Hel76,Hol11},  which subsumes the well-known observables of standard quantum mechanics. The \emph{quantum Cram\'{e}r-Rao bound} (henceforth ``q-CR bound'') \citep{Hel67,Hel68,Hel73,Hel76,Hol11} provides a lower bound on the variance of any unbiased estimator $\hat{\theta}$ of a parameter $\theta$ indexing a family of density operators $\{\rho_\theta\}$ optimized over \emph{all possible} POVMs subject to the unbiasedness condition. 

We simply state the result of the q-CR bound here, referring to \citep{Hel67,Hel68,Hel73,Hel76,Hol11} for details. We are given a family  $\{\rho_\theta\}$ of density operators depending on the parameter of interest $\theta$. The operator equation 
\begin{align} \label{lyapunov}
\frac{\partial \rho_\theta}{\partial \theta} = \frac{1}{2}\lp L_\theta \rho_\theta + \rho_\theta L_\theta\rp,
\end{align}
has a unique Hermitian solution $L_\theta = L_\theta^\dag$  when $\rho_\theta$ has no zero eigenvalues \citep{Bha07}. The operator $L_\theta$ is called the \emph{symmetric logarithmic derivative} (SLD) in analogy with the classical case. The quantity
\begin{align} \label{fisher}
I_Q(\theta) = \Tr \lp \rho_\theta \, L_\theta^2 \rp = \mean{L_\theta^2}_{\rho_\theta}
\end{align}
is known as the \emph{quantum Fisher information} and the q-CR bound reads
\begin{align} \label{qCR}
\mbox{Var}\;\hat{\theta} \geq \frac{1}{I_Q(\theta)},
\end{align}
and is valid for any unbiased estimator satisfying $\tsf{E}[\hat{\theta}] = \theta$, where the expectation is over the probability distribution induced by the POVM on the state $\rho_\theta$.

\subsection{Estimating source temperature}
The q-CR bound was originally developed in the context of quantum optical communication in the years following the invention of the laser, so it was natural for the early work to focus on the experimentally-important Gaussian states of light, particularly on the estimation of the mean vector (eqn.~\eqref{p}) of Gaussian states.  The problem of estimating the average photon number in a thermal state was also considered by \citet{Hel68}. Very recently, the estimation of a general parameter indexing single-mode and multi-mode Gaussian states has been considered by \citet{PJT+13} and \citet{Mon13} respectively.

With the density operator for filtered thermal radiation now at hand from our modal decomposition in Sec.~2.2, we may invoke the required q-CR bound from \citep{Hel68,PJT+13,Mon13}. For completeness, however, we re-derive the q-CR bound for estimating the average photon number $n_0$, to begin with, in a single-mode  thermal state $\rho_{\tsf{th}}(n_0)$. From the first representation in eq.~\eqref{thstate}, we see that
\begin{align}
\rho_{\tsf{th}}(n_0) = \frac{1}{n_0+1}\, \lp \frac{n_0}{n_0+1}\rp^N,
\end{align}
where $N = a^\dag\,a = \sum_{k=0}^\infty k \ket{k}\bra{k}$ is the number operator. Since $n_0 >0$, $\rho_{\tsf{th}}(n_0)$ has no zero eigenvalues and a unique SLD exists. To find it, we compute the derivative
\begin{align}
\frac{\partial \rho_{\tsf{th}}(n_0)}{\partial n_0} &= \frac{1}{(n_0+1)^2}\lb\frac{N}{n_0}\lp \frac{n_0}{n_0+1}\rp^N - \lp \frac{n_0}{n_0+1}\rp^N \rb \\
& = \rho_{\tsf{th}}(n_0) \lb \frac{N}{n_0(n_0+1)} - \frac{1}{n_0+1} \rb.
\end{align}
Comparing with eq.~\eqref{lyapunov}, we obtain the SLD operator
\begin{align}
L_{n_0} = \frac{N}{n_0(n_0+1)} - \frac{1}{n_0+1}.
\end{align} 
Note that  $L_{n_0}$ commutes with $\rho_{\tsf{th}}(n_0)$, reflecting the fact that the $\{\rho_{\tsf{th}}(n_0)\}$ commute with each other. The quantum Fisher information is 
\begin{align}
I_Q(n_0) &= \mean{L_{n_0}^2}_{\rho_{\tsf{th}}(n_0)} \\
&=\frac{1}{(n_0+1)^2} \lb\mean{\frac{N^2}{n_0^2}- \frac{2N}{n_0} + 1}_{\rho_{\tsf{th}}(n_0)}\rb\\
&= \frac{1}{(n_0+1)^2} \lb \frac{2\mean{N}_{\rho_{\tsf{th}}(n_0)}^2 \label{41} +\mean{N}_{\rho_{\tsf{th}}(n_0)} }{n_0^2} - \frac{2\mean{N}_{\rho_{\tsf{th}}(n_0)}}{n_0} +1 \rb\\
&= \frac{1}{n_0 (n_0+1)},
\end{align}
where we have applied Gaussian moment factoring \citep{MW95} to obtain eq.~\eqref{41} -- cf. also eqn. (25) of \citep{Hel68} and eqn.~(19) of \citep{PJT+13} for the final result.

According to eq.~\eqref{state}, the input field is the tensor product of $M= \Delta \nu T$ modes each in the thermal state $\rho_{\tsf{th}}(n_0)$. From the additivity of the quantum Fisher information for tensor-product states (which follows directly from eqs.~\eqref{lyapunov}-\eqref{fisher}), we get the total Fisher information
\begin{align}
I_Q^{\mbox{\relsize{-2}total}}(n_0) &=  \frac{\Delta \nu T}{n_0 (n_0+1)},
\end{align}
leading to the sought q-CR bound
\begin{align}\label{qCRfinal}
\mbox{Var}\; \hat{n}_0 \geq \frac{n_0(n_0+1)}{\Delta\nu T}
\end{align}
valid for any unbiased estimator $\hat{n}_0$ for $n_0$. The relative sensitivity of any unbiased estimator $\hat{n}_0$ therefore satisfies
\begin{align}\label{qCRfinal}
\frac{\mbox{Var}\; \hat{n}_0}{n_0^2} \geq \frac{n_0+1}{n_0\Delta\nu T}.
\end{align}
This lower limit on the relative sensitivity of any unbiased estimator of $n_0$ is our main result. From eq.~\eqref{psd}, we see that the relative sensitivity of an unbiased estimator $\hat{T}_s$ of the source temperature similarly obeys the limit
\begin{align} \label{qCRT}
\frac{\mbox{Var}\; \hat{T}_s}{T_s^2} \geq \frac{1+\frac{h\nu_0}{kT_s}}{\Delta\nu T}.
\end{align}

\section{Discussion}
First, we note that the estimators  $\hat{n}_0^{\mbox{\relsize{-2}(count)}}$ corresponding to photon counting and the estimator $\hat{n}_0^{\mbox{\relsize{-2}(LKD)}}$ corresponding to the two-detector scheme of \citet{LKD15} are unbiased \citep{Zmu15}, therefore the q-CR bound derived above applies to them. Eq.~\eqref{qCRfinal} agrees with the relative sensitivity of eq.~\eqref{Zmuresult} found by \citet{Zmu15} for both estimators. It also agrees with the sensitivity of eq.~\eqref{radiometer} for the ideal radiometer in the photon-rich $n_0 \gg 1$ regime. This is not surprising because in this limit, the radiometer -- which performs  a heterodyne measurement followed by post-processing to convert information on the two quadratures into a photon-number or energy measurement -- essentially counts photons \citep{Nit94,Zmu03}. However, the limit of eq.~\eqref{qCRfinal} is greater than that of eq.~\eqref{LKDresult} in the ${T_{\tsf{samp}}\ll \tau_c \ll T}$ regime for which the benefit of the two-detector scheme is claimed. As such, it provides a refutation of eq.~\eqref{LKDresult} independent of that in \citep{Zmu15}.

The q-CR limit eqs.~\eqref{qCRfinal}-\eqref{qCRT} says much more, however. Recall that the q-CR bound is applicable to all POVM measurements made on the input state subject to the unbiasedness condition. Since any concrete measurement scheme, ideal or non-ideal, corresponds mathematically to a POVM, we have shown that no possible unbiased measurement can improve on the sensitivity of  eqs.~\eqref{qCRfinal}-\eqref{qCRT}. This sensitivity is therefore ``future-proof'' and cannot be improved upon by an ingeniously designed measurement or by future technological developments. Interestingly, when $n_0 \gg 1$, the ideal (noiseless) radiometer already approaches this sensitivity limit, making it near quantum-optimal. 

\section{Acknowledgements}
This material is based on work supported by the Singapore National Research Foundation under NRF Grant No.~NRF-NRFF2011-07.

\appendix
{
\section{Calculation of mean and covariance matrix}
We calculate here the first and second-order moments $\mean{a_m}$, $\mean{a_m\,a_n}$, and $\mean{a_m^{\dag}\,a_n}$ for the filtered thermal radiation described in Section 2.1. Using eqn.~\eqref{mean}, we get $\mean{a_m} = \mean{a_m^\dag} = 0$ so that
\begin{align} \label{p}
\overline{\mb{R}}=0.
\end{align}
Similarly, eq.~\eqref{Kp} gives
\begin{align} \label{q}
\mean{a_m\, a_n} &=0.
\end{align}
Using eq.~\eqref{Kn}, we get
\begin{align}
\mean{a^{\dag}_m\,a_n} & = \frac{1}{T} \int_{-T/2}^{T/2} \int_{-T/2}^{T/2} K^{(n)}(t,t') \exp\lp -i \frac{2\pi(mt-nt')}{T}\rp\, dt\,dt' \\
&= \frac{1}{2T} \int_{-T}^{T} d\zeta \int_{-(T - \abs{\zeta})}^{T- \abs{\zeta}} d\tau K^{(n)}(\tau) \exp\lp -i \frac{2\pi}{T} \lb m \lp \frac{\zeta - \tau}{2}\rp - n \lp\frac{\zeta + \tau}{2} \rp\rb\rp \\
&=  \frac{n_0}{2\tau_cT} \int_{-T}^{T} d\zeta  \exp \lb - i \frac{2 \pi}{2T}(m-n) \zeta\rb \int_{-(T - \abs{\zeta})}^{T- \abs{\zeta}} d\tau \, \sinc \lp\frac{\tau}{\tau_c}\rp  \exp\lp -i {2\pi} \lb \nu_0 - \frac{m+n}{2T}\rb \tau\rp, \label{c}
\end{align}
where we have changed variables to $\zeta = t' +t$ and $\tau = t' -t$. We now exploit the fact that the coherence time $\tau_c \ll T$, the observation time. Consider a time duration $c \tau_c$ a few coherence times long, where $c$ is a small number, say  $c \approx 5$. We split the integral in eq.~\eqref{c} into three parts:-
\begin{align}
&\mean{a^{\dag}_m\,a_n} = \\
&  \frac{n_0}{2\tau_cT} \int_{-T}^{-(T- c \tau_c)} d\zeta  \exp \lb - i \frac{2 \pi}{2T}(m-n) \zeta\rb \int_{-(T - \abs{\zeta})}^{T- \abs{\zeta}} d\tau \, \sinc \lp\frac{\tau}{\tau_c}\rp  \exp\lp -i {2\pi} \lb \nu_0 - \frac{m+n}{2T}\rb \tau\rp \nonumber\\
&+  \frac{n_0}{2\tau_cT} \int_{-(T- c \tau_c)}^{T- c \tau_c} d\zeta  \exp \lb - i \frac{2 \pi}{2T}(m-n) \zeta\rb \int_{-(T - \abs{\zeta})}^{T- \abs{\zeta}} d\tau \, \sinc \lp\frac{\tau}{\tau_c}\rp  \exp\lp -i {2\pi} \lb \nu_0 - \frac{m+n}{2T}\rb \tau\rp \nonumber\\
&+
 \frac{n_0}{2\tau_cT} \int_{T- c \tau_c}^{T} d\zeta  \exp \lb - i \frac{2 \pi}{2T}(m-n) \zeta\rb \int_{-(T - \abs{\zeta})}^{T- \abs{\zeta}} d\tau \, \sinc \lp\frac{\tau}{\tau_c}\rp  \exp\lp -i {2\pi} \lb \nu_0 - \frac{m+n}{2T}\rb \tau\rp. \nonumber
\end{align}
Consider the first and third terms above. For any $\zeta$, the inner integrals in these terms range over an interval of size less than or equal to $2c\tau_c$.  Since, $\sinc(x) \leq 1$, the inner integral is bounded in absolute value by $2c\tau_c$. The outer integral is over a $\zeta$-range of $c\tau_c$, so each of these terms is bounded in absolute value by $n_0 c^2 \tau_c/T$. Since we are assuming $\tau_c \ll T$, we can neglect these terms in this limit. On the other hand, the inner integral in the second term ranges over a $\tau$ interval around $\tau=0$ that is equal to or greater than $c\tau_c$. Since most of the area under the sinc function is contained in the first few sidelobes, we can approximate, for $\abs{\zeta} \leq T-c\tau_c$,
\begin{align}
& \int_{-(T - \abs{\zeta})}^{T- \abs{\zeta}} d\tau \, \sinc \lp\frac{\tau}{\tau_c}\rp  \exp\lp -i {2\pi} \lb \nu_0 - \frac{m+n}{2T}\rb \tau\rp \nonumber \\
& \simeq \int_{-\infty}^{\infty} d\tau \, \sinc \lp\frac{\tau}{\tau_c}\rp  \exp\lp -i {2\pi} \lb \nu_0 - \frac{m+n}{2T}\rb \tau\rp \nonumber \\
& = \tau_c\, \mbox{rect}\lb \frac{\nu_0 - \frac{m+n}{2T}}{\Delta \nu} \rb,
\end{align}
where 
\begin{align}
\mbox{rect} (x)=
\left\{
	\begin{array}{ll}
		1  & \mbox{if }\abs{x}\leq 1/2  \\
		0 & \mbox{otherwise}
	\end{array}
\right.
\end{align}
is the rectangle function. Evaluating the outer integral, again using $\tau_c \ll T$,  gives
\begin{align}
\mean{a^{\dag}_m\,a_n} &\simeq n_0\, \mbox{rect}\lb \frac{\nu_0 - \frac{m}{T}}{\Delta \nu} \rb\cdot \delta_{mn}.
\end{align}
}
\bibliography{TempEstbibnew}

\begin{thebibliography}{30}
\expandafter\ifx\csname natexlab\endcsname\relax\def\natexlab#1{#1}\fi
\expandafter\ifx\csname bibnamefont\endcsname\relax
  \def\bibnamefont#1{#1}\fi
\expandafter\ifx\csname bibfnamefont\endcsname\relax
  \def\bibfnamefont#1{#1}\fi
\expandafter\ifx\csname citenamefont\endcsname\relax
  \def\citenamefont#1{#1}\fi
\expandafter\ifx\csname url\endcsname\relax
  \def\url#1{\texttt{#1}}\fi
\expandafter\ifx\csname urlprefix\endcsname\relax\def\urlprefix{URL }\fi
\providecommand{\bibinfo}[2]{#2}
\providecommand{\eprint}[2][]{\url{#2}}

\bibitem[{\citenamefont{Lieu et~al.}(2015)\citenamefont{Lieu, Kibble, and
  Duan}}]{LKD15}
\bibinfo{author}{\bibfnamefont{R.}~\bibnamefont{Lieu}},
  \bibinfo{author}{\bibfnamefont{T.~W.~B.} \bibnamefont{Kibble}},
  \bibnamefont{and} \bibinfo{author}{\bibfnamefont{L.}~\bibnamefont{Duan}},
  \bibinfo{journal}{The Astrophysical Journal} \textbf{\bibinfo{volume}{798}},
  \bibinfo{pages}{67} (\bibinfo{year}{2015}),
  \urlprefix\url{http://stacks.iop.org/0004-637X/798/i=2/a=67}.

\bibitem[{\citenamefont{Dicke}(1946)}]{Dic46}
\bibinfo{author}{\bibfnamefont{R.~H.} \bibnamefont{Dicke}},
  \bibinfo{journal}{Review of Scientific Instruments}
  \textbf{\bibinfo{volume}{17}}, \bibinfo{pages}{268} (\bibinfo{year}{1946}).

\bibitem[{\citenamefont{Burke and Graham-Smith}(2010)}]{BG-S10}
\bibinfo{author}{\bibfnamefont{B.~F.} \bibnamefont{Burke}} \bibnamefont{and}
  \bibinfo{author}{\bibfnamefont{F.}~\bibnamefont{Graham-Smith}},
  \emph{\bibinfo{title}{An introduction to radio astronomy}}
  (\bibinfo{publisher}{Cambridge University Press}, \bibinfo{year}{2010}).

\bibitem[{\citenamefont{Hanbury~Brown and Twiss}(1957)}]{HBT57}
\bibinfo{author}{\bibfnamefont{R.}~\bibnamefont{Hanbury~Brown}}
  \bibnamefont{and} \bibinfo{author}{\bibfnamefont{R.~Q.} \bibnamefont{Twiss}},
  \bibinfo{journal}{Proceedings of the Royal Society of London. Series A.
  Mathematical and Physical Sciences} \textbf{\bibinfo{volume}{242}},
  \bibinfo{pages}{300} (\bibinfo{year}{1957}).

\bibitem[{\citenamefont{Zmuidzinas}(2015)}]{Zmu15}
\bibinfo{author}{\bibfnamefont{J.}~\bibnamefont{Zmuidzinas}},
  \bibinfo{journal}{arXiv preprint arXiv:1501.03219}  (\bibinfo{year}{2015}).

\bibitem[{\citenamefont{Nityananda}({1994})}]{Nit94}
\bibinfo{author}{\bibfnamefont{R.}~\bibnamefont{Nityananda}}, in
  \emph{\bibinfo{booktitle}{{Very High Angular Resolution Imaging}}}, edited by
  \bibinfo{editor}{\bibnamefont{{Robertson, JG and Tango, WJ}}}
  (\bibinfo{year}{{1994}}), no. \bibinfo{number}{{158}} in
  \bibinfo{series}{{IAU Symposia}}, pp. \bibinfo{pages}{{11--18}},
  \bibinfo{note}{{158th Symposium of the International Astronomical Union}}.

\bibitem[{\citenamefont{Zmuidzinas}(2003{\natexlab{a}})}]{Zmu03}
\bibinfo{author}{\bibfnamefont{J.}~\bibnamefont{Zmuidzinas}},
  \bibinfo{journal}{Applied optics} \textbf{\bibinfo{volume}{42}},
  \bibinfo{pages}{4989} (\bibinfo{year}{2003}{\natexlab{a}}).

\bibitem[{\citenamefont{Helstrom}(1976)}]{Hel76}
\bibinfo{author}{\bibfnamefont{C.~W.} \bibnamefont{Helstrom}},
  \emph{\bibinfo{title}{Quantum Detection and Estimation Theory}}
  (\bibinfo{publisher}{Academic Press}, \bibinfo{year}{1976}).

\bibitem[{\citenamefont{Holevo}(2011)}]{Hol11}
\bibinfo{author}{\bibfnamefont{A.~S.} \bibnamefont{Holevo}},
  \emph{\bibinfo{title}{Probabilistic and Statistical Aspects of Quantum
  Theory}} (\bibinfo{publisher}{Edizioni Della Normale, Pisa, Italy},
  \bibinfo{year}{2011}).

\bibitem[{\citenamefont{Giovannetti et~al.}(2011)\citenamefont{Giovannetti,
  Lloyd, and Maccone}}]{GLM11}
\bibinfo{author}{\bibfnamefont{V.}~\bibnamefont{Giovannetti}},
  \bibinfo{author}{\bibfnamefont{S.}~\bibnamefont{Lloyd}}, \bibnamefont{and}
  \bibinfo{author}{\bibfnamefont{L.}~\bibnamefont{Maccone}},
  \bibinfo{journal}{Nature Photonics} \textbf{\bibinfo{volume}{5}},
  \bibinfo{pages}{222} (\bibinfo{year}{2011}).

\bibitem[{\citenamefont{Nielsen and Chuang}(2000)}]{NC00}
\bibinfo{author}{\bibfnamefont{M.~A.} \bibnamefont{Nielsen}} \bibnamefont{and}
  \bibinfo{author}{\bibfnamefont{I.~L.} \bibnamefont{Chuang}},
  \emph{\bibinfo{title}{Quantum Computation and Quantum Information}}
  (\bibinfo{publisher}{Cambridge University Press}, \bibinfo{year}{2000}).

\bibitem[{\citenamefont{Marzolino and Braun}(2013)}]{MB13}
\bibinfo{author}{\bibfnamefont{U.}~\bibnamefont{Marzolino}} \bibnamefont{and}
  \bibinfo{author}{\bibfnamefont{D.}~\bibnamefont{Braun}},
  \bibinfo{journal}{Phys. Rev. A} \textbf{\bibinfo{volume}{88}},
  \bibinfo{pages}{063609} (\bibinfo{year}{2013}),
  \urlprefix\url{http://link.aps.org/doi/10.1103/PhysRevA.88.063609}.

\bibitem[{\citenamefont{Marzolino and Braun}(2015)}]{MB15}
\bibinfo{author}{\bibfnamefont{U.}~\bibnamefont{Marzolino}} \bibnamefont{and}
  \bibinfo{author}{\bibfnamefont{D.}~\bibnamefont{Braun}},
  \bibinfo{journal}{Phys. Rev. A} \textbf{\bibinfo{volume}{91}},
  \bibinfo{pages}{039902} (\bibinfo{year}{2015}),
  \urlprefix\url{http://link.aps.org/doi/10.1103/PhysRevA.91.039902}.

\bibitem[{\citenamefont{Jarzyna and Zwierz}(2014)}]{JZ14}
\bibinfo{author}{\bibfnamefont{M.}~\bibnamefont{Jarzyna}} \bibnamefont{and}
  \bibinfo{author}{\bibfnamefont{M.}~\bibnamefont{Zwierz}},
  \bibinfo{journal}{arXiv preprint arXiv:1412.56099}  (\bibinfo{year}{2014}).

\bibitem[{\citenamefont{Shapiro}(2009)}]{Sha09}
\bibinfo{author}{\bibfnamefont{J.~H.} \bibnamefont{Shapiro}},
  \bibinfo{journal}{Selected Topics in Quantum Electronics, IEEE Journal of}
  \textbf{\bibinfo{volume}{15}}, \bibinfo{pages}{1547 } (\bibinfo{year}{2009}),
  ISSN \bibinfo{issn}{1077-260X}.

\bibitem[{\citenamefont{Mandel and Wolf}(1995)}]{MW95}
\bibinfo{author}{\bibfnamefont{L.}~\bibnamefont{Mandel}} \bibnamefont{and}
  \bibinfo{author}{\bibfnamefont{E.}~\bibnamefont{Wolf}},
  \emph{\bibinfo{title}{Optical coherence and quantum optics}}
  (\bibinfo{publisher}{Cambridge University Press}, \bibinfo{year}{1995}).

\bibitem[{\citenamefont{Mandel}(1959)}]{Man59}
\bibinfo{author}{\bibfnamefont{L.}~\bibnamefont{Mandel}},
  \bibinfo{journal}{Proceedings of the Physical Society}
  \textbf{\bibinfo{volume}{74}}, \bibinfo{pages}{233} (\bibinfo{year}{1959}).

\bibitem[{\citenamefont{Weedbrook et~al.}(2012)\citenamefont{Weedbrook,
  Pirandola, Garc\'ia-Patr\'on, Cerf, Ralph, Shapiro, and Lloyd}}]{WPG+12}
\bibinfo{author}{\bibfnamefont{C.}~\bibnamefont{Weedbrook}},
  \bibinfo{author}{\bibfnamefont{S.}~\bibnamefont{Pirandola}},
  \bibinfo{author}{\bibfnamefont{R.}~\bibnamefont{Garc\'ia-Patr\'on}},
  \bibinfo{author}{\bibfnamefont{N.~J.} \bibnamefont{Cerf}},
  \bibinfo{author}{\bibfnamefont{T.~C.} \bibnamefont{Ralph}},
  \bibinfo{author}{\bibfnamefont{J.~H.} \bibnamefont{Shapiro}},
  \bibnamefont{and} \bibinfo{author}{\bibfnamefont{S.}~\bibnamefont{Lloyd}},
  \bibinfo{journal}{Rev. Mod. Phys.} \textbf{\bibinfo{volume}{84}},
  \bibinfo{pages}{621} (\bibinfo{year}{2012}).

\bibitem[{\citenamefont{Olivares}(2012)}]{Oli12}
\bibinfo{author}{\bibfnamefont{S.}~\bibnamefont{Olivares}},
  \bibinfo{journal}{The European Physical Journal-Special Topics}
  \textbf{\bibinfo{volume}{203}}, \bibinfo{pages}{3} (\bibinfo{year}{2012}).

\bibitem[{\citenamefont{Cram{\'e}r}(1946)}]{Cra46}
\bibinfo{author}{\bibfnamefont{H.}~\bibnamefont{Cram{\'e}r}},
  \emph{\bibinfo{title}{Mathematical methods of statistics}}
  (\bibinfo{publisher}{Princeton University Press}, \bibinfo{year}{1946}).

\bibitem[{\citenamefont{Rao}(1945)}]{Rao45}
\bibinfo{author}{\bibfnamefont{C.~R.} \bibnamefont{Rao}},
  \bibinfo{journal}{Bulletin of the Calcutta Mathematical Society}
  \textbf{\bibinfo{volume}{37}}, \bibinfo{pages}{81} (\bibinfo{year}{1945}).

\bibitem[{\citenamefont{{Van Trees}}(2001)}]{VanTreesI}
\bibinfo{author}{\bibfnamefont{H.~L.} \bibnamefont{{Van Trees}}},
  \emph{\bibinfo{title}{Detection, Estimation, and Modulation Theory : Part I}}
  (\bibinfo{publisher}{Wiley-Interscience 1st Ed}, \bibinfo{year}{2001}),
  \bibinfo{edition}{1st} ed.

\bibitem[{\citenamefont{Zmuidzinas}(2003{\natexlab{b}})}]{Zmu03b}
\bibinfo{author}{\bibfnamefont{J.}~\bibnamefont{Zmuidzinas}},
  \bibinfo{journal}{JOSA A-Optics, Image Science, and Vision}
  \textbf{\bibinfo{volume}{20}}, \bibinfo{pages}{218}
  (\bibinfo{year}{2003}{\natexlab{b}}).

\bibitem[{\citenamefont{Yadav et~al.}(2007)\citenamefont{Yadav, Komatsu, and
  Wandelt}}]{YKW07}
\bibinfo{author}{\bibfnamefont{A.~P.~S.} \bibnamefont{Yadav}},
  \bibinfo{author}{\bibfnamefont{E.}~\bibnamefont{Komatsu}}, \bibnamefont{and}
  \bibinfo{author}{\bibfnamefont{B.~D.} \bibnamefont{Wandelt}},
  \bibinfo{journal}{The Astrophysical Journal} \textbf{\bibinfo{volume}{664}},
  \bibinfo{pages}{680} (\bibinfo{year}{2007}).

\bibitem[{\citenamefont{Helstrom}(1967)}]{Hel67}
\bibinfo{author}{\bibfnamefont{C.~W.} \bibnamefont{Helstrom}},
  \bibinfo{journal}{Physics Letters A} \textbf{\bibinfo{volume}{25}},
  \bibinfo{pages}{101} (\bibinfo{year}{1967}).

\bibitem[{\citenamefont{Helstrom}(1968)}]{Hel68}
\bibinfo{author}{\bibfnamefont{C.~W.} \bibnamefont{Helstrom}},
  \bibinfo{journal}{Information Theory, IEEE Transactions on}
  \textbf{\bibinfo{volume}{14}}, \bibinfo{pages}{234} (\bibinfo{year}{1968}).

\bibitem[{\citenamefont{Helstrom}(1973)}]{Hel73}
\bibinfo{author}{\bibfnamefont{C.~W.} \bibnamefont{Helstrom}},
  \bibinfo{journal}{International Journal of Theoretical Physics}
  \textbf{\bibinfo{volume}{8}}, \bibinfo{pages}{361} (\bibinfo{year}{1973}).

\bibitem[{\citenamefont{Bhatia}(2007)}]{Bha07}
\bibinfo{author}{\bibfnamefont{R.}~\bibnamefont{Bhatia}},
  \emph{\bibinfo{title}{Positive definite matrices}}
  (\bibinfo{publisher}{Princeton University Press}, \bibinfo{year}{2007}).

\bibitem[{\citenamefont{Pinel et~al.}(2013)\citenamefont{Pinel, Jian, Treps,
  Fabre, and Braun}}]{PJT+13}
\bibinfo{author}{\bibfnamefont{O.}~\bibnamefont{Pinel}},
  \bibinfo{author}{\bibfnamefont{P.}~\bibnamefont{Jian}},
  \bibinfo{author}{\bibfnamefont{N.}~\bibnamefont{Treps}},
  \bibinfo{author}{\bibfnamefont{C.}~\bibnamefont{Fabre}}, \bibnamefont{and}
  \bibinfo{author}{\bibfnamefont{D.}~\bibnamefont{Braun}},
  \bibinfo{journal}{Physical Review A} \textbf{\bibinfo{volume}{88}},
  \bibinfo{pages}{040102} (\bibinfo{year}{2013}).

\bibitem[{\citenamefont{Monras}(2013)}]{Mon13}
\bibinfo{author}{\bibfnamefont{A.}~\bibnamefont{Monras}},
  \bibinfo{journal}{arXiv preprint arXiv:1303.3682}  (\bibinfo{year}{2013}).

\end{thebibliography}
\bibliographystyle{apsrev}

\end{document}